\documentclass[conference]{IEEEtran}

\usepackage[cmex10]{amsmath}
\usepackage{amssymb}

\usepackage{graphicx}
\usepackage{cite}
\ifCLASSINFOpdf
\else
\fi
\newtheorem{defi}{Definition}
\newtheorem{theorem}{Theorem}
\newtheorem{lemma}{Lemma}

\newtheorem{corollary}{Corollary}

\begin{document}
%
\title{Optimal Power and Rate Allocation in the Degraded Gaussian Relay Channel with Energy Harvesting Nodes}

\author{\IEEEauthorblockN{Mahmood Mohassel Feghhi, Aliazam Abbasfar, \emph{Senior Member}, \emph{IEEE}, Mahtab Mirmohseni}
\IEEEauthorblockA{School of Electrical and Computer Eng., College of Eng., University of Tehran\\
Tehran 14395-515, IRAN\\
Emails: mohasselfeghhi@ut.ac.ir, abbasfar@ut.ac.ir, m.mirmohseni@ece.ut.ac.ir}
}


%


\maketitle

\begin{abstract}
Energy Harvesting (EH) is a novel technique to prolong the lifetime of the wireless networks such as wireless sensor networks or Ad-Hoc networks, by providing an unlimited source of energy for their nodes. In this sense, it has emerged as a promising technique for Green Communications, recently. On the other hand, cooperative communication with the help of relay nodes improves the performance of wireless communication networks by increasing the system throughput or the reliability as well as the range and efficient energy utilization. In order to investigate the cooperation in EH nodes, in this paper, we consider the problem of optimal power and rate allocation in the degraded full-duplex Gaussian relay channel in which source and relay can harvest energy from their environments. We consider the general stochastic energy arrivals at the source and the relay with known EH times and amounts at the transmitters before the start of transmission. This problem has a min-max optimization form that along with the constraints is not easy to solve. We propose a method based on a mathematical theorem proposed by Terkelsen \cite{terk} to transform it to a solvable convex optimization form. Also, we consider some special cases for the harvesting profile of the source and the relay nodes and find their solutions efficiently.
\end{abstract}


%
\IEEEpeerreviewmaketitle

\section{Introduction}
\begin{figure*}[!ht]
\centering
\includegraphics[width=.8\linewidth]{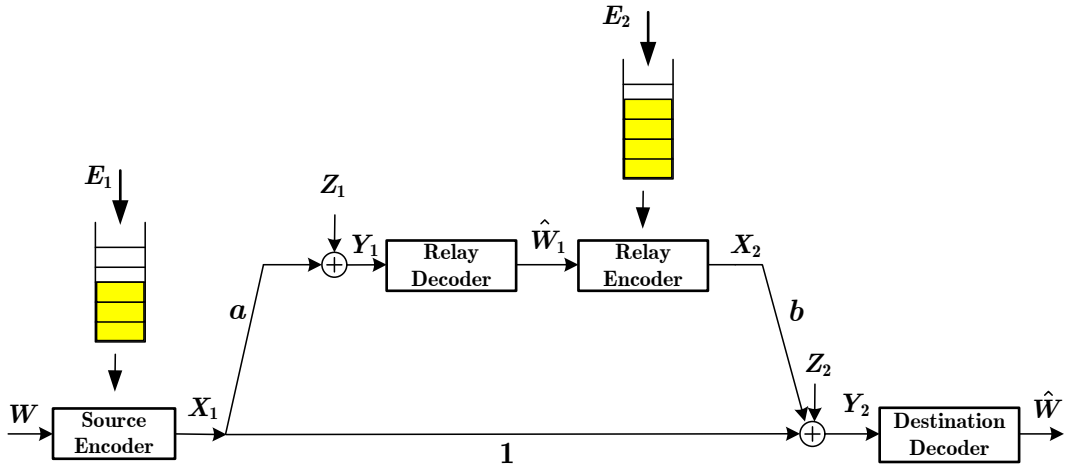}
\caption{Gaussian relay channel with energy harvesting nodes.}
\label{fig:1}
\end{figure*}
Energy Harvesting (EH) has emerged as a promising solution to the perennial energy constraint of wireless networks such as wireless sensor networks (WSN) or Ad-hoc networks, which have limited battery sources \cite{sudev}. Also, EH is developed to be used as a foundation of green communication networks \cite{zhu}. This will be more critical as increasing energy consumption of highly-demanded ubiquitous networks soon be the main cause of global warming. Energy harvesters collect ambient energy from the environment or the resources such as body heat and convert it into usable electrical energy. Conventional devices that harvest energy from the environment are solar cells, water mills, wind turbines, microbial fuel cells, vibration absorption devices, thermo-electric generators, piezoelectric cells, etc. EH nodes have access to an unlimited source of energy in contrast to conventional battery-powered nodes; however, the limitation on the EH production rate and its sporadic nature necessitates the sophisticated utilization of scavenged energy.
 
Early works on EH sensor nodes are presented in \cite{raghun, kansal, sharma}. In \cite{rajesh}, shannon capacity of EH sensor nodes, which sends information over an AWGN channel is studied. Also, the effects of energy buffer with different condition and also inefficiencies in energy storage are studied. Similar study to \cite{rajesh} is presented in \cite{ozel} which showed that AWGN channel capacity with stochastic energy arrival (i.e., EH) is the same as AWGN channel capacity with average power constraint equals to average recharge rate. Also, two achievability schemes are presented, namely, save-and-transmit policy and best-effort-transmit policy. Optimal packet scheduling problem in wireless single user EH communication system, in which energy and data packets are stochastically entered the source node is considered in \cite{yang}. In the aforementioned paper, in order to minimize the transmit time of the data packets, transmission rate adaptively changes according to data and energy traffics. The study in \cite{yang} is then extended to broadcast channel \cite{yang_ozel, antepli}, multiple-access channel \cite{yang_jcn}, interference channel \cite{tutun}, two-hop network \cite{gunduz} and fading channels \cite{ozel_jsac}.  

Wireless Relay Channel (RC), ever since introduced, is used to overcome the challenges of information transmission in wireless channels with progressively improving protocols proposed by researchers. Cooperative communications based on the use of wireless relay nodes is a specific area of research in wireless communications that extensively explored in the last decade, from many aspects such as information theoretic capacity, diversity, outage analysis, cooperative and network coding, resource allocation, etc. Also, resource-constraint networks such as WSN can get benefit of cooperation through optimal allocation of energy and bandwidth to the nodes based on the available channel state information of those nodes (see e.g. \cite{hong} and the references therein).

Some recent studies have considered multi-hop and relay networks where their nodes are capable of harvesting ambient energy \cite{gunduz},\cite{orhan_CISS, huang}. In EH two-hop network considered in \cite{gunduz}, only the relay node can harvest energy, while in two-hop networks studied in \cite{orhan_CISS} both the source (S) and the relay (R) nodes are the EH nodes. In \cite{huang}, half-duplex (HD) orthogonal RC with decode-and-forward (DF) relay is considered and two different delay constraints, namely, one-block decoding delay constraint and arbitrary decoding delay constraint up to the total transmission blocks, are investigated.

 In this paper, we consider the problem of optimal power and rate allocation for EH nodes of a three-node full-duplex (FD) degraded Gaussian RC in order to maximize the total number of bits that can be delivered from the source node to the destination (D) node in a given deadline. We consider a general model compared to that of the aforementioned papers. In our model, there exists a direct link from S to D (in contrast to \cite{gunduz},\cite{orhan_CISS}) and also we investigate the FD mode compared to the HD mode of \cite{huang}. This causes a more complicated min-max optimization problem arises in our scenario which has not been encountered in the previous studies. Our aim is to transform this complicated min-max problem to a solvable convex optimization form, using some mathematical background. In two steps, we first introduce an auxiliary parameter and then use a minimax theorem of \cite{terk} to make our problem tractable. Since the online problem that assigns rate and power in real-time to the nodes, in our studied scenario, is intractable for now, we consider the offline solution that the time instants of energy harvesting and the amount of harvested energy by S and R are known before the course of the transmission. Furthermore, some special cases on the harvesting profile of S and R are investigated. The first and the second cases have only one node (S or R), which harvests energy from its environment, and in the third case, harvesting profiles of S and R are the same. These special cases are presented to give some intuition of the main problem.

The remainder of the paper is organized as follows. Section \ref{sec:sysmodel} introduces the system model,  and section \ref{sec:formulation} formulates the throughput maximization problem for the degraded Gaussian RC. In section \ref{sec:optimalsol}, we provide the optimal solution for the degraded Gaussian RC and in section \ref{sec:specialcase}, we investigate some special cases. Finally, section \ref{sec:coclusion} concludes the paper. 
\section{System Model}\label{sec:sysmodel}
RC models a three-node network, in which the source node wants to communicate to the destination node with the help of the relay node.

\begin{defi}\label{def:code}
A $({2^{nR}},n)$ code for the discrete memoryless RC (DM-RC) $({{\cal X}_1} \times {{\cal X}_2},p({y_2},{y_1}|{x_2},{x_1}),{{\cal Y}_1} \times {{\cal Y}_2})$  with four finite sets ${{\cal X}_1},{{\cal X}_2},{{\cal Y}_1},{{\cal Y}_2}$ and conditional probability mass functions $p({y_2},{y_1}|{x_2},{x_1})$ on $ {{\cal Y}_1} \times {{\cal Y}_2} $ consists of (i) A message set $ \left[ {1:{2^{nR}}} \right] $, (ii) An encoder that assigns a codeword $ x_1^n(m) $ to each message $ m \in \left[ {1:{2^{nR}}} \right] $, (iii) A R encoder that assigns a symbol $ {x_{2i}}(y_1^{i - 1}) $ to each past received sequence $ y_1^{i - 1} \in {\cal Y}_1^{i - 1} $ for each $ i \in \left[ {1:n} \right] $, and (iv) A decoder that assigns an estimate $ \tilde m $ to each received sequence $ y_2^n \in {\cal Y}_2^n $ or reports an error. The channel is said to be memoryless in the sense that given the current transmitted symbols $ ({X_{1i}},{X_{2i}}) $, the current received symbols $ ({Y_{1i}},{Y_{2i}}) $ are conditionally independent of the message and the past transmitted and received symbols $ (m,{X_1^{i - 1}},X_2^{i - 1},Y_1^{i - 1},{Y_2^{i - 1}}) $. We assume uniform distribution of the message over its set. The average probability of error is defined as $ P_e^{(n)} = Pr\{ \tilde M \ne M\}  $. 
\end{defi}

\begin{defi}\label{def:rate}
A rate $ {\cal R} $ is said to be achievable for the DM-RC if there exists a sequence of $ ({2^{nR}},n) $ codes such that $ {\lim _{n \to \infty }}P_e^{(n)} = 0 $. The capacity $ C $ of the DM-RC is the supremum of all achievable rates. 
\end{defi} 

We consider Gaussian RC with EH nodes depicted in Fig.~ \ref{fig:1}. 
The channel outputs correspond to channel input $ {X_1},{X_2} $ is as follows
\begin{IEEEeqnarray}{rCl}
{Y_1} & = & a{X_1} + {Z_1},\label{y_relay}\\
{Y_2} & = & {X_1} + b{X_2} + {Z_2},\label{y_destination}
\end{IEEEeqnarray}
where $ a $ and $ b $ are channel gains of S-R and R-D links, respectively, assuming normalized channel gain for the S-D link, and we have $ {Z_1} \sim {\cal N}(0,N),\,\,{Z_2} \sim {\cal N}(0,N) $.

Capacity of the degraded RC is as follows \cite{cover}
\begin{equation}
C = \mathop {\max }\limits_{p({x_1},{x_2})} \min \left\{ {I({X_1},{X_2};{Y_2}),I({X_1};{Y_1}|{X_2})} \right\},
\end{equation}
and the capacity formula for the degraded Gaussian RC with power constraint at S, $ \sum\nolimits_{i = 1}^n {x_{1i}^2(w) \le n{P_1}}  $, and R, $ \sum\nolimits_{i = 1}^n {x_{2i}^2(w) \le n{P_2}}  $, is given by \cite{cover}

$\!\!\!\!\!\! C({P_1},{P_2}) =\min \{ {{\tilde C}_1},{{\tilde C}_2}\}= $
\begin{equation}
\left\{
\begin{IEEEeqnarraybox}[\IEEEeqnarraystrutmode
\IEEEeqnarraystrutsizeadd{2pt}{2pt}][c]{lll}
{{{\tilde C}_1} = {C\left( {\frac{{{{\left( {\sqrt {{P_1}({a^2}{P_1} - {b^2}{P_2})}  + \sqrt {{b^2}({a^2} - 1){P_1}{P_2}} } \right)}^2}}}{{{a^2}{P_1}N}}} \right),}}  \\ \qquad\qquad\qquad\qquad\qquad\qquad\qquad\qquad \textrm{if} \;\;\; {\frac{{({a^2} - 1){P_1}}}{{{P_2}}} \ge 1},\\
{{{\tilde C}_2} = C\left( {\frac{{\max \left\{ {1,{a^2}} \right\}{P_1}}}{N}} \right),}\qquad\qquad\qquad \qquad \textrm{otherwise},
\end{IEEEeqnarraybox}
\right.\label{gaussiancap}
\end{equation}
which is achieved by $ {X_1} \sim {\cal N}(0,{P_1}) $ and $ {X_2} \sim {\cal N}(0,{P_2}) $.
\section{Problem Formulation}\label{sec:formulation}
\begin{figure}[tb]
\centering
\includegraphics[width=1\linewidth]{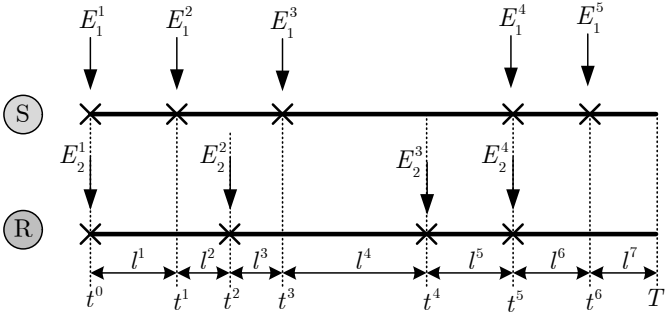}
\caption{EH instants and amounts for S and R with $ K=6 $.}
\label{fig:2}
\end{figure}

Our problem is to maximize the number of bits delivered by a deadline $ T $ from S to D. S and R harvest energy at random instants $ {t^0},{t^1},{t^2},...,{t^K} $ and in random amounts $ E_1^1,E_1^2,...,E_1^{K+1} $ and $ E_2^1,E_2^2,...,E_2^{K+1} $, respectively. If at some instants only S or R harvests energy, we simply set the amounts of the energy harvested by the other one to zero (see Fig. \ref{fig:2}).
The interval between two instants S or R or both harvest energy from the environment, will be called an epoch. The length of $ i^{th} $  epoch is  $ {l^i} = {t^i} - {t^{i - 1}} $ for $ i = 1,...,K+1 $. So, there are a total of $ K+1 $ epoch with $ {t^0}=0 $ and $ {l^{K + 1}} = T - {t^K} $. We consider the offline problem in which the arrival times and amounts are known to S and R before the start of transmission; therefore, $ {l^i},\,\,\,i = 1,...,K+1 $ are known ahead of time. We find optimal power allocation for S and R in order to maximize the rate from S to D with energy causality constraints at S and R. This means that energy cannot be utilized in S or R before it is harvested in the corresponding node. We formulate the problem as follows:
\begin{equation}
\!\!\mathop {\max }\limits_{{P_1},{P_2}} \sum\limits_{i = 1}^{K+1} {\min } \left\{ {\tilde C_1^{}(P_1^i,P_2^i),\tilde C_2^{}(P_1^i,P_2^i)} \right\}\! = \!\mathop {\max }\limits_{{P_1},{P_2}} \sum\limits_{i = 1}^{K+1} {C(P_1^i,P_2^i)} 
\end{equation}
\begin{IEEEeqnarray}{rll}
s.t.\qquad& P_1^i \ge 0,\,P_2^i \ge 0,&\qquad i = 1,...,K + 1,\label{nonneg}\\
& \sum\limits_{i = 1}^k {P_1^i{l^i} \le \sum\limits_{i = 0}^{k - 1} {E_1^i} } ,&\qquad k = 1,...,K + 1,\label{Scausal}\\
&\sum\limits_{i = 1}^k {P_2^i{l^i} \le \sum\limits_{i = 0}^{k - 1} {E_2^i} } ,&\qquad k = 1,...,K + 1.\label{Rcausal}
\end{IEEEeqnarray}

Equation (\ref{nonneg}) denotes that the powers of S and R should be nonnegative, (\ref{Scausal}) states the energy causality at S and (\ref{Rcausal}) states that power consumption at R should not violate its energy causality constraint. 
Finding the solution of the main problem is not straightforward as it has the min-max optimization form that cannot be separated due to the FD nature of the problem. In other word, since R can receive and send the information at the same time, in each epoch we do not know which term (in epoch $ i $, $\tilde C_1^{}(P_1^i,P_2^i) $ or $ \tilde C_2^{}(P_1^i,P_2^i) $)  is the minimum and how should we assign the powers of S and R to maximize the sum of the rates allocated to all of the epochs. Also, observe that in (\ref{gaussiancap}) the condition that specifies the minimum term, depends on the optimization parameters of the maximization problem, i.e., $ {P_1} $ and $ {P_2} $.

\section{Optimal Solution for Degraded Gaussian RC}\label{sec:optimalsol}
In this section, we propose a method to make the problem tractable and try to solve it. We can rewrite the problem by introducing $ 0 \le \lambda  \le 1 $ as follows
\begin{equation}
\mathop {\max }\limits_{\{ P_{_1}^i\} ,\{ P_2^i\}} \sum\limits_{i = 1}^{K+1} {\mathop {\min }\limits_{\{ {\lambda ^i}\} }  } \,\left\{ {{\lambda ^i}\tilde C_1^{}(P_1^i,P_2^i) + (1 - {\lambda ^i})\tilde C_2^{}(P_1^i,P_2^i)} \right\}\label{opt_prob}
\end{equation}
\begin{IEEEeqnarray}{rll}
s.t. \qquad& P_1^i \ge 0,\,P_2^i \ge 0,& \qquad i = 1,...,K + 1,\\
&\sum\limits_{i = 1}^k {P_1^i{l^i} \le \sum\limits_{i = 0}^{k - 1} {E_1^i} },&\qquad k = 1,...,K + 1,\\
&\sum\limits_{i = 1}^k {P_1^i{l^i} \le \sum\limits_{i = 0}^{k - 1} {E_1^i} },&\qquad k = 1,...,K + 1,
\end{IEEEeqnarray}
where,
\begin{equation}
{\lambda ^i} = 
\begin{cases}
0& \textrm{if}\qquad \tilde C_1^{}(P_1^i,P_2^i) > \tilde C_2^{}(P_1^i,P_2^i),\\
1&\textrm{if}\qquad \tilde C_1^{}(P_1^i,P_2^i) < \tilde C_2^{}(P_1^i,P_2^i),\\
\textrm{arbitrary}& \textrm{if}\qquad \tilde C_1^{}(P_1^i,P_2^i) = \tilde C_2^{}(P_1^i,P_2^i).
\end{cases}
\end{equation}
We then use the following corollary to change the order of min and max operators which is the application of a min-max theorem of Terkelsen \cite{terk}, presented in \cite{geng}.

\newcounter{tempequationcounter}
\begin{figure*}[!t]
\normalsize
\setcounter{tempequationcounter}{\value{equation}}
\begin{IEEEeqnarray*}{rCl}
\setcounter{equation}{30}
{\cal L}(\{ P_1^i\} ,\{ P_2^i\} ,\xi ,\mu ,\vartheta ,\eta ) &=& \sum\limits_{i = 1}^{K + 1} {\left\{ {{\lambda ^i}C\left( {\frac{{{{\left( {\sqrt {P_1^i({a^2}P_1^i - {b^2}P_2^i)}  + \sqrt {{b^2}({a^2} - 1)P_1^iP_2^i} } \right)}^2}}}{{{a^2}P_1^iN}}} \right) + (1 - {\lambda ^i})C\left( {\frac{{\max \left\{ {1,{a^2}} \right\}P_1^i}}{N}} \right)} \right\}\,{l^i}}\\
&-& \sum\limits_{k = 1}^K  {{\xi _k}\left( {\sum\limits_{i = 1}^k {P_1^i{l^i}}  - \sum\limits_{i = 0}^{k - 1} {E_1^i} } \right)}  - \sum\limits_{k = 1}^K {{\mu _k}\left( {\sum\limits_{i = 1}^k {P_2^i{l^i}}  - \sum\limits_{i = 0}^{k - 1} {E_2^i} } \right)}  + \sum\limits_{i = 1}^{K + 1} {{\vartheta _i}P_1^i}  + \sum\limits_{i = 1}^{K + 1} {{\eta _i}P_2^i} ,\yesnumber\label{longequation}
\end{IEEEeqnarray*}
\setcounter{equation}{\value{tempequationcounter}}
\hrulefill
\vspace{4pt}
\end{figure*}
\begin{corollary}[{\cite[Corollary 1]{geng}}]\label{corollary1}
Let $ {\Lambda _d} $ be the d-dimensional simplex, i.e. $ {\lambda _i} \ge 0 $  and $ \,\sum\nolimits_{i = 1}^d {{\lambda _i} = 1} $ . Let $ {\cal P} $ be the set of probability distributions $ p(u) $. Let $ {T_i}(p(u)),\,i = 1,...,d $ be a set of functions such that the set $ {\cal A} $,
\begin{equation}
{\cal A}\! =\! \left\{\! {({a_1},{a_2},...,{a_d}) \in {\mathbb{R} ^d}:{a_i} \le {T_i}(p(u))\,\textrm{for}\,\textrm{some}\,\,p(u) \in {\cal P}}\! \right\}\!,
\end{equation} 
is a convex set. Then,
\begin{equation}
\mathop {\sup }\limits_{p(u) \in {\cal P}} \mathop {\min }\limits_{\lambda  \in {\Lambda _d}} \sum\limits_{i = 1}^d {{\lambda _i}{T_i}(p(u)) = } \mathop {\min }\limits_{\lambda  \in {\Lambda _d}} \mathop {\sup }\limits_{p(u) \in {\cal P}} \sum\limits_{i = 1}^d {{\lambda _i}{T_i}(p(u))}.
\end{equation}
\end{corollary}

Now, we consider our optimization problem defined in (\ref{opt_prob}).

\begin{theorem}\label{theorem1}
Optimal power and rate allocation for the source and the relay node in the degraded Gaussian RC with energy harvesting nodes, is the solution of the following problem
\begin{equation}
\mathop {\min }\limits_{\{ {\lambda ^i}\} } \mathop {\max }\limits_{\{ P_{_1}^i\} ,\{ P_2^i\} } \sum\limits_{i = 1}^{K+1} {\,\left\{ {{\lambda ^i}\tilde C_1^{}(P_1^i,P_2^i) + (1 - {\lambda ^i})\tilde C_2^{}(P_1^i,P_2^i)} \right\}}
\end{equation}
\begin{IEEEeqnarray}{rll}
s.t.\qquad& P_1^i \ge 0,\,P_2^i \ge 0,&\qquad i = 1,...,K + 1,\\
&\sum\limits_{i = 1}^k {P_1^i{l^i} \le \sum\limits_{i = 0}^{k - 1} {E_1^i} },&\qquad k = 1,...,K + 1,\\
&\sum\limits_{i = 1}^k {P_2^i{l^i} \le \sum\limits_{i = 0}^{k - 1} {E_2^i} },&\qquad k = 1,...,K + 1,\\
& 0 \le {\lambda ^i} \le 1,&\qquad i = 1,...,K + 1.
\end{IEEEeqnarray}
\end{theorem}

\begin{IEEEproof}
We show that the corollary \ref{corollary1} is applicable to our problem. In other words, we have
\begin{equation}
\begin{array}{l}
\mathop {\max }\limits_{\{ P_{_1}^i\} ,\{ P_2^i\} } \sum\limits_{i = 1}^{K+1} {\mathop {\min }\limits_{\{ {\lambda ^i}\} } } \,\left\{ {{\lambda ^i}\tilde C_1^{}(P_1^i,P_2^i) + (1 - {\lambda ^i})\tilde C_2^{}(P_1^i,P_2^i)} \right\} = \\
\,\,\,\,\,\,\,\,\,\,\mathop {\min }\limits_{\{ {\lambda ^i}\} } \mathop {\max }\limits_{\{ P_{_1}^i\} ,\{ P_2^i\} } \sum\limits_{i = 1}^{K+1} {\,\left\{ {{\lambda ^i}\tilde C_1^{}(P_1^i,P_2^i) + (1 - {\lambda ^i})\tilde C_2^{}(P_1^i,P_2^i)} \right\}}.\label{main_prob} 
\end{array}
\end{equation}

Now, we show that every convex combination of the points inside $ {\cal A} $ is also in $ {\cal A} $. Therefore, $ {\cal A} $ is a convex set. Consider the following mutual information terms 
\begin{IEEEeqnarray*}{rCl}
I({X_1};{Y_1}|{X_2},Q)&\mathop  = \limits^{(a)}& H({Y_1}|{X_2},Q) - H({Y_1}|{X_1},{X_2},Q)\\
&\mathop  \le \limits^{(b)}& H({Y_1}|{X_2}) - H({Y_1}|{X_1},{X_2},Q)\\
&\mathop  = \limits^{(c)}& H({Y_1}|{X_2}) - H({Y_1}|{X_1},{X_2})\\
&\mathop  = \limits^{(d)}& I({X_1};{Y_1}|{X_2}),\yesnumber
\end{IEEEeqnarray*}
\begin{IEEEeqnarray*}{rCl}
I({X_1},{X_2};{Y_2}|Q)&\mathop  = \limits^{(a)}& H({Y_2}|Q) - H({Y_2}|{X_1},{X_2},Q)\\
&\mathop  \le \limits^{(b)}& H({Y_2}) - H({Y_2}|{X_1},{X_2},Q)\\
&\mathop  = \limits^{(c)} & H({Y_2}) - H({Y_2}|{X_1},{X_2})\\
&\mathop  = \limits^{(d)}& I({X_1},{X_2};{Y_2}),\yesnumber
\end{IEEEeqnarray*}
where, (a) and (d) follow from the definition of the mutual information, (b) follows from the fact that conditioning does not increase the entropy, and (c) follows from the fact that $ Q $ is a function of $ {X_1} $ and $ {X_2} $. This completes the proof.

Also, we can specially show that the aforementioned corollary is applicable to the Gaussian case. We set $ {T_1}(p(u)) = \tilde C_1^{}(P_1^{},P_2^{}) $, $ {T_2}(p(u)) = \tilde C_2^{}(P_1^{},P_2^{}) $ and $ d=2 $.
Note that as we fix the input distribution of S and R to be Gaussian, i.e., $ {X_1} \sim {\cal N}(0,{P_1}) $ and $ {X_2} \sim {\cal N}(0,{P_2}) $, we can replace $ {T_i}(p({x_1},{x_2})) $ with $ {T_i}({P_1},{P_2}) $ in the above corollary (we rewrite $ p({x_1},{x_2}) = g({P_1},{P_2}) $).
Now, assume that $ ({a_1},{a_2}) \in {\cal A}\, $ and $ ({b_1},{b_2}) \in {\cal A} $. It means that $ {a_1} \le C_1^0(P_1^{},P_2^{}) $,$ {a_2} \le C_2^0(P_1^{},P_2^{}) $ as well as $ {b_1} \le C_1^0(P_1^{},P_2^{}) $, $ {b_2} \le C_2^0(P_1^{},P_2^{}) $. Then we choose $ ({c_1},{c_2}) = (\eta {a_1} + (1 - \eta ){b_1},\eta {a_2} + (1 - \eta ){b_2}),\,\,\,0 \le \eta  \le 1 $. It is clear that $ {c_1} \le C_1^0(P_1^{},P_2^{}) $ and $ {c_2} \le C_2^0(P_1^{},P_2^{}) $. Hence, we have $ ({c_1},{c_2}) \in {\cal A} $. This completes the proof.\\
\end{IEEEproof}

Now, we decompose our problem into the following two problems.

\begin{corollary}\label{corollary2}
The problem defined in (\ref{main_prob}) can be decomposed into the following problems:
\begin{equation}
\begin{array}{l}
{\!\!\!\!\!({\rm{Problem}}\,{\rm{1}}):{f^*}(\{ {\lambda ^i}\} )\! = \!\mathop {\max }\limits_{\{ P_{_1}^i\} ,\{ P_2^i\} } \sum\limits_{i = 1}^{K + 1}\! {\left\{ {{\lambda ^i}\tilde C_1^{}(P_1^i,P_2^i)} \right.}} \\{
\,\,\,\,\,\,\,\,\,\,\,\;\;\;\;\,\,\,\qquad\qquad\,\,\,\,\,\,\,\,\,\,\,\,\,\,\,\,\,\,\,\,\,\,\,\,\,\,\,\,\,\,\,\,\,\,\,\,\,\,\,\,\,\,\,\,\,\,\,\left. { + (1 - {\lambda ^i})\tilde C_2^{}(P_1^i,P_2^i)} \right\}}
\end{array}
\end{equation}
\begin{IEEEeqnarray}{rll}
s.t.\qquad& P_1^i \ge 0,\,P_2^i \ge 0,&\qquad i = 1,...,K + 1,\\
&\sum\limits_{i = 1}^k {P_1^i{l^i} \le \sum\limits_{i = 0}^{k - 1} {E_1^i} },&\qquad k = 1,...,K + 1,\\
&\sum\limits_{i = 1}^k {P_2^i{l^i} \le \sum\limits_{i = 0}^{k - 1} {E_2^i} },&\qquad k = 1,...,K + 1.
\end{IEEEeqnarray}
\begin{IEEEeqnarray}{lll}
\!\!\!\!\!\!\!\!\!\!\!\!\!\!({\rm{Problem}}\,2):&\,\,\mathop {\min }\limits_{\{ {\lambda ^i}\} } \,\,\,{f^*}(\{ {\lambda ^i}\} )& \label{problem2}\\
& s.t.\,\,\,\,\,\,\,\,0 \le {\lambda ^i} \le 1,\,\,\,\,\,\,\,\,\,\,i = 1,...,K + 1.&
\end{IEEEeqnarray}
which can be solved, separately. Problem 1 is a convex optimization problem as its objective function is concave and its constraints are affine, and can be solved by efficient convex optimization methods to find its unique maximizer. Problem 2 is a combinatorial problem that can be solved efficiently, too.
\end{corollary}
For the first problem, we can write the Lagrangian function for any $ {\xi _k} \ge 0,\,\,{\mu _k} \ge 0,{\vartheta _k} \ge 0$ and ${\eta _k} \ge 0 $ as (\ref{longequation}) in top of this page together with following complementary slackness conditions
\addtocounter{equation}{1}
\begin{IEEEeqnarray}{rl}
{\xi _k}\left( {\sum\limits_{i = 1}^k {P_1^i{l^i}}  - \sum\limits_{i = 0}^{k - 1} {E_1^i} } \right) = 0, &\;\;\; k = 1,...,K, \label{slackness}\\
{\mu _k}\left( {\sum\limits_{i = 1}^k {P_2^i{l^i}}  - \sum\limits_{i = 0}^{k - 1} {E_2^i} } \right) = 0,&\;\;\; k = 1,...,K,\\
\sum\limits_{i = 1}^N {{\vartheta _i}P_1^i}  = 0,&\;\;\; i = 1,...,K + 1,\\
\sum\limits_{i = 1}^N {{\eta _i}P_2^i}  = 0,&\;\;\; i = 1,...,K + 1.
\end{IEEEeqnarray}

This problem can be solved by taking the derivatives of the Lagrangian function with respect to $ {P_1} $  and $ {P_2} $ and setting them to zero, and doing some mathematical manipulation; however, the closed form expression for $ {P_1} $  and $ {P_2} $ give not any explicit idea about the optimal power assignment algorithm. Hence, in the next we prove some lemmas about the properties of the optimal solution. We use these lemmas in the next section to find the optimal solution for a special case.

\begin{lemma}
In an optimal policy, transmit rates and powers of the S and R are constant within an energy harvesting epoch and only potentially change at energy harvesting instants.
\end{lemma}

\begin{IEEEproof}
As we know, $ {\tilde C_1} = {g_1}({P_1},{P_2}) $, $ {\tilde C_2} = {g_2}({P_1}) $ are nonnegative, strictly concave and monotonically increasing function of their variables $ {P_1} $  and $ {P_2} $. We prove this lemma by contradiction. Assume that there is a $ {t^*} \in ({t^{j - 1}},{t^j}) $  such that S and R use $ {\hat P_1} $  and $ {\hat P_2} $ in $ ({t^{j - 1}},{t^*}) $ and $ {\breve P_1} $  and $ {\breve P_2} $ in $ ({t^*},{t^j}) $, respectively. Hence we have
\begin{eqnarray}
\lefteqn{\frac{{{t^*} - {t^{j - 1}}}}{{{t^j} - {t^{j - 1}}}}{g_1}({{\hat P}_1},{{\hat P}_2}) + \frac{{{t^j} - {t^*}}}{{{t^j} - {t^{j - 1}}}}{g_1}({{\breve P}_1},{{\breve P}_2}) \le}\nonumber\\
&\!\! {g_1}\left(\!\! {\frac{{({t^*} - {t^{j - 1}}){{\hat P}_1} + ({t^j} - {t^*}){{\breve P}_1}}}{{{t^j} - {t^{j - 1}}}},\frac{{({t^*} - {t^{j - 1}}){{\hat P}_2} + ({t^j} - {t^*}){{\breve P}_2}}}{{{t^j} - {t^{j - 1}}}}}\!\! \right).
\end{eqnarray}
Similarly
\begin{eqnarray}
\lefteqn{\frac{{{t^*} - {t^{j - 1}}}}{{{t^j} - {t^{j - 1}}}}{g_2}({\hat P_1}) + \frac{{{t^j} - {t^*}}}{{{t^j} - {t^{j - 1}}}}{g_2}({\breve P_1}) \le}\nonumber\\
 & {g_2}\left( {\frac{{({t^*} - {t^{j - 1}}){{\hat P}_1} + ({t^j} - {t^*}){{\breve P}_1}}}{{{t^j} - {t^{j - 1}}}}} \right).
\end{eqnarray}
Therefore by equalizing the transmitted power within an epoch we can reach to a higher throughput. Hence, changing the transmitted power of S and R within an epoch is suboptimal.
\end{IEEEproof}

\begin{lemma}
Whenever the power of source or relay changes, it should only increase.
\end{lemma}

\begin{IEEEproof}
This is also due to the concavity of $ {\tilde C_1} = {g_1}({P_1},{P_2}) $, $ {\tilde C_2} = {g_2}({P_1}) $ and the fact that postponing the transmission of energy or shifting it to the right (in energy consumption diagram) does not violate the energy causality constraint and on the other hand, due to the concavity of the rate function in terms of power, more bits per joule can be sent by setting the power to a constant value. Therefore, the power of S or R never decreases in time, i.e., $ P_1^1 \le P_1^2 \le P_1^3 \le  \cdots  $ and $ P_2^1 \le P_2^2 \le P_2^3 \le  \cdots  $.
\end{IEEEproof}

\begin{corollary}
In the optimal policy, if power of S or R changes in an instant, the total harvested energy in the previous epoch of that node has been consumed completely by this instant.
\end{corollary}

\section{Some Special Cases}\label{sec:specialcase}
In this section we present some special cases, which are interesting from the practical viewpoint. We can consider cases in which only one node, i.e. S or R, can harvest energy from the environment.
Also there may be scenarios that the harvesting process of S and R are the same. These special cases do not have the complexity of the main problem and presented here to give intuition to the main problem. Although three cases are solvable, we only state the solution for the third one, which is general compared to the others.
  
\subsection{Only Relay Harvests Energy}
In this scenario, the topology of the network is such that only R can harvest energy from its environment and S has solely a non-replenishable battery. This is equivalent to the case that in our system model we set $ E_1^1 \ne 0,E_1^2 = ... = E_1^K = 0 $. In this case we have $ P_1^i = \frac{{E_1^1}}{T} = P,\,\forall i$ and the capacity formula is as follows
\begin{equation}
{\hat C_I}({P_2}) = \left\{ {\begin{array}{*{20}{c}}
{\!\!\!\!\!\!\!\!\!\!\!\!\!\!\!\!\!\!\!\!\!\!\!\!\!\!\!\!\!\!\!\!\!\!\!C\left( {\frac{{{{\left( {\sqrt {{a^2}P - {b^2}{P_2}}  + b\sqrt {({a^2} - 1){P_2}} } \right)}^2}}}{{{a^2}N}}} \right),}\\{\qquad\qquad\qquad\qquad\qquad\qquad \textrm{if}\,\,{P_2} \le \,({a^2} - 1)P}\\
{\!\!\!C\left( {\frac{{\max \left\{ {1,{a^2}} \right\}P}}{N}} \right),\qquad\qquad\qquad\qquad\qquad \textrm{o.w.}}
\end{array}} \right.\label{nonEHsource}
\end{equation}

As we can see in (\ref{nonEHsource}), the condition that specifies the capacity formula is only the function of $ {P_2} $ and therefore the complexity of the main problem is not exists here; so, finding the solution of this problem is straightforward. 
\subsection{Only Source Harvests Energy}
This scenario is in contrast to that of the previous one, in which only S can harvest ambient energy while R has a conventional non-rechargeable battery. This means that in our model we set $ E_2^1 \ne 0,E_2^2 = ... = E_2^K = 0 $; So, $ P_2^i = \frac{{E_2^1}}{T} = P,\,\forall i $ and the capacity in this scenario is as
\begin{equation}
{\hat C_{II}}({P_1}) = \left\{ {\begin{array}{*{20}{c}}
{\!\!\!\!\!\!\!\!\!\!\!\!\!\!\!\!\!\!\!\!\!\!\!\!\!\!\!\!\!\!\!\!\!\!\!C\left( {\frac{{{{\left( {\sqrt {{a^2}{P_1} - {b^2}P}  + b\sqrt {({a^2} - 1)P} } \right)}^2}}}{{{a^2}N}}} \right),}\\{\qquad\qquad\qquad\qquad\qquad\qquad \textrm{if}\,\,{P_1} \ge \frac{P}{{\,({a^2} - 1)}}}\\
{\!\!\!C\left( {\frac{{\max \left\{ {1,{a^2}} \right\}{P_1}}}{N}} \right),\qquad\qquad\qquad\qquad\qquad \textrm{o.w.}}
\end{array}} \right.\label{nonEHrelay}
\end{equation}
The same conclusion can be made as the previous case.

\subsection{Same Harvesting Process for Source and Relay}
This is also an interesting case that both of S and R can harvest the ambient energy. In this scenario, S and R are considered in the vicinity of each other; therefore, we can assume the same harvesting profile for them, i.e., the harvesting instants are the same and the harvested amounts are scaled version of each other. Mathematically speaking, we have $ E_1^i = \gamma E_2^i,\,\,\,i = 1,...,K, $ and thus $ P_2^i = \gamma P_1^i,\,\forall i $, for some positive constant $ \gamma $. The capacity formula is given as
\begin{equation}
{\hat C_{III}}({P_1}) = \left\{ {\begin{array}{*{20}{c}}
{\!\!\!\!\!\!\!\!\!\!\!\!\!\!\!\!\!\!\!\!\!\!\!\!\!\!\!\!\!\!\!\!C\left( {\frac{{{{\left( {\sqrt {{a^2} - {b^2}\gamma }  + b\sqrt {\gamma ({a^2} - 1)} } \right)}^2}{P_1}}}{{{a^2}N}}} \right),\,\,\,\,\,\,}\\{\qquad\qquad\qquad if \,\,\frac{{\,({a^2} - 1)}}{\gamma } \ge 1\,\,or\,\,\,a \ge \sqrt {\gamma  + 1} }\\
{\!\!\!\!\!\!\!C\left( {\frac{{\max \left\{ {1,{a^2}} \right\}{P_1}}}{N}} \right),\,\,\,\,\,\,\,\,\,\,\,\,\,\,\,\qquad\,\,\,\,\,\,\,\,\,\,\,\,\,\,\,\,\,\,\,\,\,\,\,\,o.w.}
\end{array}} \right.\label{capacity_example3}
\end{equation}
We obtain, the Lagrangian of our problem in this case as
\begin{IEEEeqnarray*}{lll}
{\cal L}(\{ P_1^i\} ,\{ P_2^i\} ,\xi ,\mu ,\vartheta ,\eta ) &=& \sum\limits_{i = 1}^{K+1} {\min } \left\{ {C_1^0(P_1^i,P_2^i),C_2^0(P_1^i,P_2^i)} \right\}{l^i}\\
& -& \sum\limits_{k = 1}^{K} {{\xi _k}\left( {\sum\limits_{i = 1}^k {P_1^i{l^i}}  - \sum\limits_{i = 0}^{k - 1} {E_1^i} } \right)} \\
& - &\sum\limits_{k = 1}^{K} {{\mu _k}\left( {\sum\limits_{i = 1}^k {P_2^i{l^i}}  - \sum\limits_{i = 0}^{k - 1} {E_2^i} } \right)} \\
& + & \sum\limits_{i = 1}^{K+1} {{\vartheta ^i}P_1^i}  + \sum\limits_{i = 1}^{K+1} {{\eta ^i}P_2^i} . \yesnumber
\end{IEEEeqnarray*}
By setting the derivative of the Lagrangian, with respect to S power sequence, to zero, we get the optimal power sequence of S as
\begin{equation}
P_1^i = \left\{ {\begin{array}{*{20}{c}}
{{{\left[ {\frac{1}{{2{A_i}}} - \frac{1}{{{K_1}}}} \right]}^ + },\,\,\,\,\,\,\,\,a \ge \sqrt {\gamma  + 1} },\\
{{{\left[ {\frac{1}{{2{A_i}}} - \frac{1}{{{K_2}}}} \right]}^ + },\,\,\,\,\,\,\,\,a < \sqrt {\gamma  + 1} },
\end{array}} \right.
\label{increase}
\end{equation}
where $ {A_i} = \sum\limits_{k = i}^K {{\xi _k} - \frac{{{\vartheta ^i}}}{{{l^i}}}}  $, 
$ {K_1} = \frac{{{{\left( {\sqrt {{a^2} - {b^2}\gamma }  + b\sqrt {\gamma ({a^2} - 1)} } \right)}^2}}}{{{a^2}N}}$ and $ {K_2} = {{\max \{ 1,{a^2}\} } \mathord{\left/
 {\vphantom {{\max \{ 1,{a^2}\} } N}} \right.
 \kern-\nulldelimiterspace} N}
 $.

As we have assumed that $ E_1^1 > 0 $ and $ E_2^1 > 0 $, therefore $ P_1^1 > 0 $ and $ P_2^1 > 0 $. Also, according to Lemma 2, we conclude that $ P_1^i > 0,\,\,\,\forall i \in \{ 1,...,K+1\}  $. This together with complementary slackness means that we have $ {\vartheta ^i} = 0,\,\,\,\forall i \in \{ 1,...,K+1\}  $, and hence $ {A_i} = \sum\nolimits_{k = i}^K {{\xi _k}}  $. Note that according to (\ref{increase}) we have $ P_1^1 \le P_1^2 \le \cdots  $ and hence $ P_2^1 \le P_2^2 \le \cdots  $. This proves Lamma 2 for this problem. 
Now, we present the following lemma for this problem.

\begin{lemma}\label{lamma3}
In the optimal policy, the power at S or R changes only when their corresponding energy causality constraints are active.
\end{lemma}

\begin{IEEEproof}
Here we give a proof for this lemma based on complementary slackness condition. 
Observe that if $ P_1^i \ne P_1^{i + 1} $ then, according to (\ref{increase}), we should have $ {\xi _i} \ne 0 $. Then, according to the complementary slackness condition of (\ref{slackness}), we should have $ \sum\nolimits_{j = 1}^i {P_1^j{l^j}}  - \sum\nolimits_{j = 0}^{i - 1} {E_1^j}  = 0 $. This completes the proof of lemma \ref{lamma3} for this problem. 
\end{IEEEproof}

Now, we can present the optimal solution form as follows:
\begin{equation}
{o_v} = \arg \,\mathop {\min }\limits_{{o_{v - 1}} < i \le k} \frac{{\sum\nolimits_{j = {o_{v - 1}}}^i {E_1^j} }}{{{t^i} - {t^{{o_{v - 1}}}}}},
\end{equation}
\begin{equation}
P_1^v = \frac{{\sum\nolimits_{j = {o_{v - 1}}}^{{o_v} - 1} {E_1^j} }}{{{t^{{o_v}}} - {t^{{o_{v - 1}}}}}},\,P_2^v = \gamma P_1^v.
\end{equation}
Once the powers of S and R are determined, their corresponding rates can be obtained using capacity formula (\ref{capacity_example3}).

\section{Conclusion}\label{sec:coclusion}
In this paper, we investigated the optimal power and rate allocation for a three-node full-duplex degraded Gaussian relay channel with energy harvesting source and relay nodes. The original problem has a complicated min-max form that is not easy to solve. We transformed it to a tractable convex optimization problem, which can be solved efficiently. Also, some special cases on the harvesting profile of the source and the relay nodes were considered.






%
\bibliographystyle{IEEEtran}
\bibliography{My_Refs}
%
%

\end{document}